\documentclass[10pt,twocolums]{iopart}
\usepackage{epsf}

\begin{document}

\title{Localization of superconductivity in superconductor-electromagnet hybrids}

\author{G W Ataklti$^{1}$,  A Yu Aladyshkin$^{1,2}$, W Gillijns$^{1}$, I M Nefedov$^{2}$, J Van de
Vondel$^{1}$, A V Silhanek$^{1,3}$, M Kemmler$^{4}$, R
Kleiner$^{4}$, D Koelle$^{4}$, V V Moshchalkov$^{1}$}

\address{$^{(1)}$ INPAC -- Institute for Nanoscale Physics and Chemistry, Nanoscale
Superconductivity and Magnetism  and Pulsed Fields Group,
K.U. Leuven, Celestijnenlaan 200D, B--3001 Leuven, Belgium \\
$^{(2)}$ Institute for Physics of Microstructures, Russian Academy
of Sciences, 603950, Nizhny Novgorod, GSP-105, Russia\\
$^{(3)}$ D\'{e}partement de Physique, Universit\'{e} de Li\`{e}ge,
B\^{a}t. B5, All\'{e}e du 6 ao\^{u}t 17, B -- 4000 Sart Tilman, Belgium \\
$^{(4)}$ Physikalisches Institut -- Experimentalphysik II and
Center for Collective Quantum Phenomena in LISA$^{+}$,
Universit\"{a}t T\"{u}bingen, Auf der Morgenstelle 14, 72076
T\"{u}bingen, Germany}



\begin{abstract}
We investigate the nucleation of superconductivity in a
superconducting Al strip under the influence of the magnetic field
generated by a current-carrying Nb wire, perpendicularly oriented
and located underneath the strip. The inhomogeneous magnetic
field, induced by the Nb wire, produces a spatial modulation of
the critical temperature $T_c$, leading to a controllable
localization of the superconducting order parameter (OP) wave
function. We demonstrate that close to the phase boundary
$T_{c}(B_{ext})$ the localized OP solution can be displaced
reversibly by either applying an external perpendicular magnetic
field $B_{ext}$ or by changing the amplitude of the inhomogeneous
field.
\end{abstract}

\maketitle

The sensitivity of superconductivity to the local strength of a
magnetic field has been exploited during the last years to confine
superconductivity by applying a non-uniform magnetic field ${\bf
b}({\bf r})$
\cite{Lange-PRL-03,Aladyshkin-PRB-03,Gillijns-PRB-07}. The
experimental realization of this ``magnetic" confinement can be
achieved, e.g., in hybrid superconductor (S) -- ferromagnet (F)
structures and ferromagnetic superconductors. The properties of
the ferromagnetic superconductors and the S/F hybrids with rather
strong exchange interaction between superconducting and
ferromagnetic subsystems were discussed in the reviews
\cite{Bulaevskii-AdvPhys-85,Izyumov-UFN-02,Buzdin-RMP-05,Bergeret-RMP-05}.
Hereafter we will focus on the flux-coupled hybrids, where the
interaction between superconducting element and the sources of the
magnetic field (e.g., domain walls in the ferromagnetic film)
occurs via slowly decaying stray fields only
\cite{Lyuksyutov-AdvPhys-05,Velez-JMMM-08,Aladyshkin-SUST-09}.


In general, for thin-film superconducting samples, infinite in the
lateral directions, superconducting order parameter (OP) wave
function first nucleates near the $|B_z(x,y)|$ minima, where
$B_z(x,y)=B_{ext}+b_z(x,y)$ is the out-of-plane component of the
total magnetic field, $B_{ext}$ is the applied external magnetic
field (see arguments, e.g., in \cite{Aladyshkin-SUST-09}).
Depending on $B_{ext}$, favorable conditions for the appearance of
superconductivity can be fulfilled either above domain walls in a
thick ferromagnetic substrate (domain--wall superconductivity
\cite{Buzdin-PRB-03,Yang-Nature-09,Werner-PRB-11}), or above
magnetic domains of opposite polarity with respect to the
$B_{ext}$ sign (reverse-domain superconductivity
\cite{Yang-PRB-06,Yang-APL-06,Fritzsche-PRL-06,Aladyshkin-APL-09,Aladyshkin-PRB-11}).
The external-field-induced crossover between domain--wall
superconductivity and reverse-domain superconductivity as
$|B_{ext}|$ increases can result in an unusual dependence of the
superconducting critical temperature $T_c$ on $B_{ext}$, which can
be nonlinear or even non-monotonous
\cite{Aladyshkin-PRB-11,Aladyshkin-PRB-06,Aladyshkin-JAP-10,Yang-APL-11}
in contrast to a plain superconducting film in a uniform magnetic
field. In addition, these domain patterns are periodic in space
and therefore superconductivity has to be located at all
magnetically compensated areas. In order to have a singly
connected OP solution a non-periodic field profile is needed
\cite{Milosevic-APL-10}. It is interesting to note that the stray
field of a single domain wall in a {\it thin} ferromagnetic layer
cannot provide domain-wall superconductivity and non-monotonous
(or, in the other words, reentrant) phase boundary $T_c(B_{ext})$
due to vanishing of the field at large distances from the domain
wall \cite{Aladyshkin-PRB-03}. Even if the amplitude of the stray
field produced by domain structure becomes insufficient to
localize superconductivity, such parallel magnetic domains can
induce preferential vortex motion and giant anisotropy of the
critical currents in superconducting films and crystals
\cite{VVlasov-PRB-08a,VVlasov-PRB-08b,Zhu-PRL-08,Ozmetin-APL-09,Belkin-APL-08,Belkin-APL-10}.
    \begin{figure}[h!]
    \begin{center}
    \epsfxsize=90mm \epsfbox{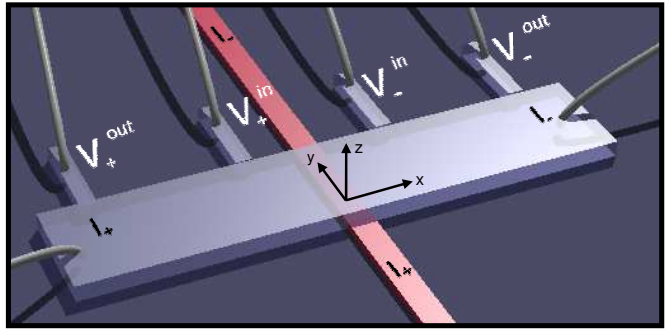} \caption{(Color online)
    Schematic representation of the sample: the Al strip (top gray
    element) placed perpendicular on top of the Nb wire (bottom red
    element). \label{Fig1-new}}
    \end{center}
    \end{figure}

We would like to note that the amplitude of the stray magnetic
field and its profile in real S/F bilayers are dictated by the
saturated magnetization of the ferromagnet and by the period of
the domain structure (or magnetic dot array), therefore the
flexibility of S/F hybrids is limited
\cite{Gillijns-PRB-07,Aladyshkin-JAP-10,Gillijns-PRL-05,Gillijns-PRB-06,Silhanek-PRB-07}.
Full control over the amplitude of the inhomogeneous magnetic
field can be reached for the hybrid structures, where the
ferromagnetic subsystem is replaced by current-carrying
coils/wires. Such superconductor-electromagnet (S/Em) hybrid
systems (also called cryotrons) were invented in 1950's and
originally considered as superconducting computer elements and
circuits controlled by local magnetic field of the coils/wires
(see classical textbooks and reviews
\cite{deGennes,Bremer,Lock,Newhouse}). However the most transport
measurements done on cryotrons were carried out at low
temperatures, when the ability to manipulate the intensive
superconducting currents seems to be the most effective. To the
best of our knowledge the report of the S/Em properties at high
temperatures was carried out by Pannetier {\it et al.}
\cite{Pannetier-95}. In this work for the S/Em system, consisting
of a plain Al film and a lithographically defined array of
parallel metallic lines, it was experimentally demonstrated that
(i) the $T_c(B_{ext})$ dependence can be non-monotonous for
considerably large driving current $I_0$ in these metallic lines
and (ii) the shape of the $T_c(B_{ext})$ can be reversibly changed
as $I_0$ varies.

In this paper we study the influence of the non-periodic magnetic
field ${\bf b}_w$, generated by a single current-carrying wire, on
the nucleation of superconductivity. The simplicity of this system
allows us to fully understand the combined influence of the
external homogeneous field and the inhomogeneous field by the
electromagnet in the migration of the superconducting order
parameter. To avoid heating effects, this current-carrying wire
was fabricated from a superconducting material (Nb) with a
considerably higher critical temperature $T_c$ than the
investigated microbridge (Al strip). Due to the design of the
sample, the perpendicular $z-$component of the magnetic field,
playing an important role for thin-film structures, is uniform
across the Al strip ($y-$axis) and it varies only along the strip
($x-$axis), vanishing slowly as the distance from the wire
increases. This configuration allows us to directly detect the
localization of the OP wave function in experiment as the
perpendicularly oriented external field $B_{ext}$ or the magnitude
of the non-uniform field \mbox{$B_0={\rm max}\,|b_{w,z}(x,y)|$}
are changed. We also show that for low magnetic fields, although
the nucleation of superconductivity occurs at those positions,
where the $z-$component of the total magnetic field is close to
zero, there is still a systematic decrease in $T_c$ as a function
of $B_{ext}$, resulting from the increasing gradient of the
magnetic field at this position of the OP localization. This work
expands our previous investigation of the vortex dynamics in a
similar system \cite{Cryotron-PRB-11} by describing the influence
of the non-uniform field of the wire on the phase boundary
$T_c(B_{ext})$.

The hybrid samples consist of a 4$\,\mu$m wide and 100 nm thick Al
strip, patterned by electron-beam lithography and lift-off
technique, placed perpendicularly on top of a 1.5$\,\mu$m wide and
50 nm thick Nb wire processed by e-beam lithography and Ar ion
milling (Fig.~\ref{Fig1-new}). In order to avoid electrical
contact, the Al strip and the current-carrying Nb wire are
separated by a 120 nm thick insulating Ge layer. All details of
the fabrication processes are given in
Ref.~\cite{Cryotron-PRB-11}. To investigate the spatial
localization of superconductivity in the Al strip, two sets of
voltage contacts were prepared at distances of 10$\,\mu$m and 50
$\,\mu$m from the Nb wire (the inner and outer contacts,
respectively).

\begin{figure}[h!]
    \begin{center}
    \epsfxsize=95mm \epsfbox{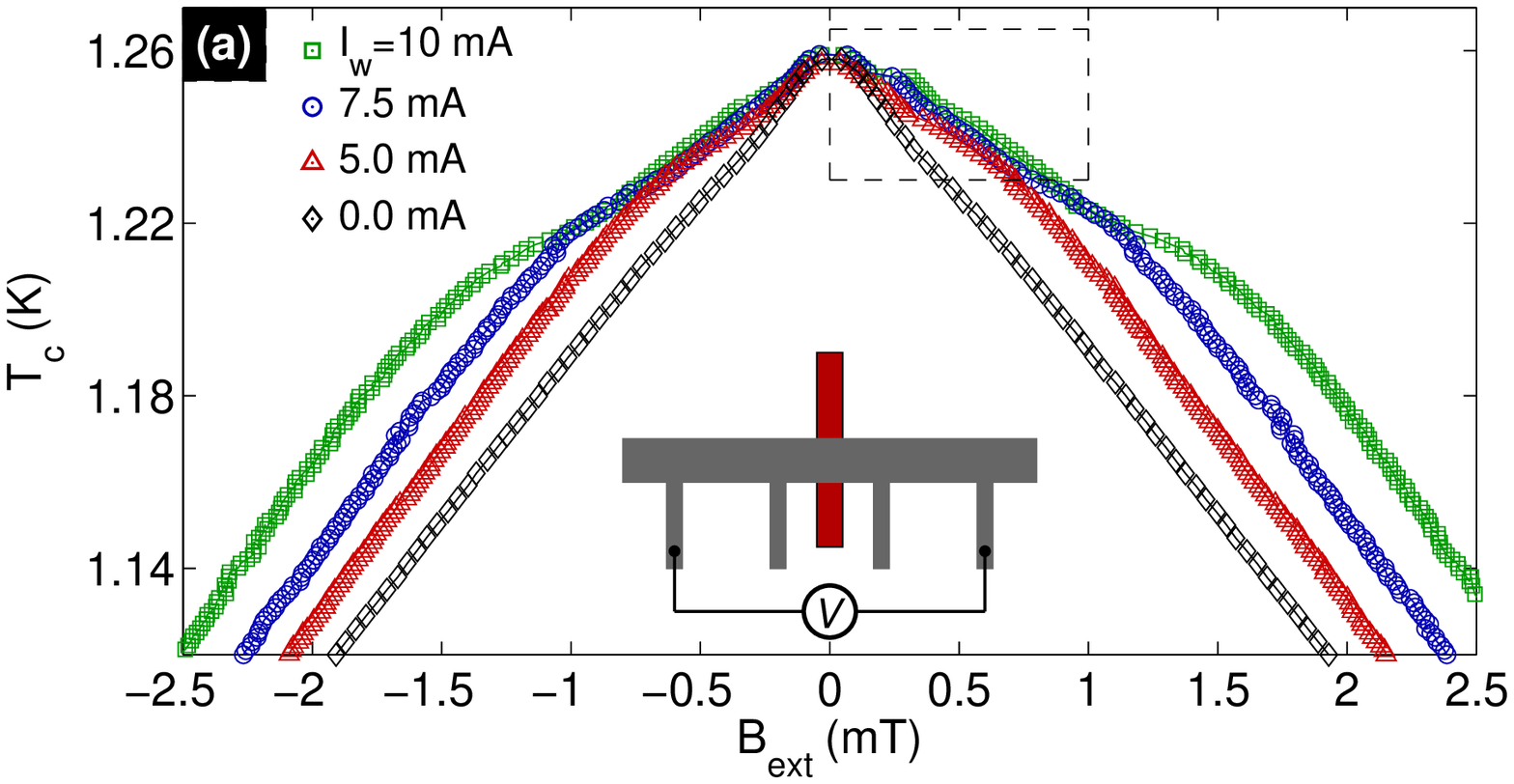}
    \epsfxsize=95mm \epsfbox{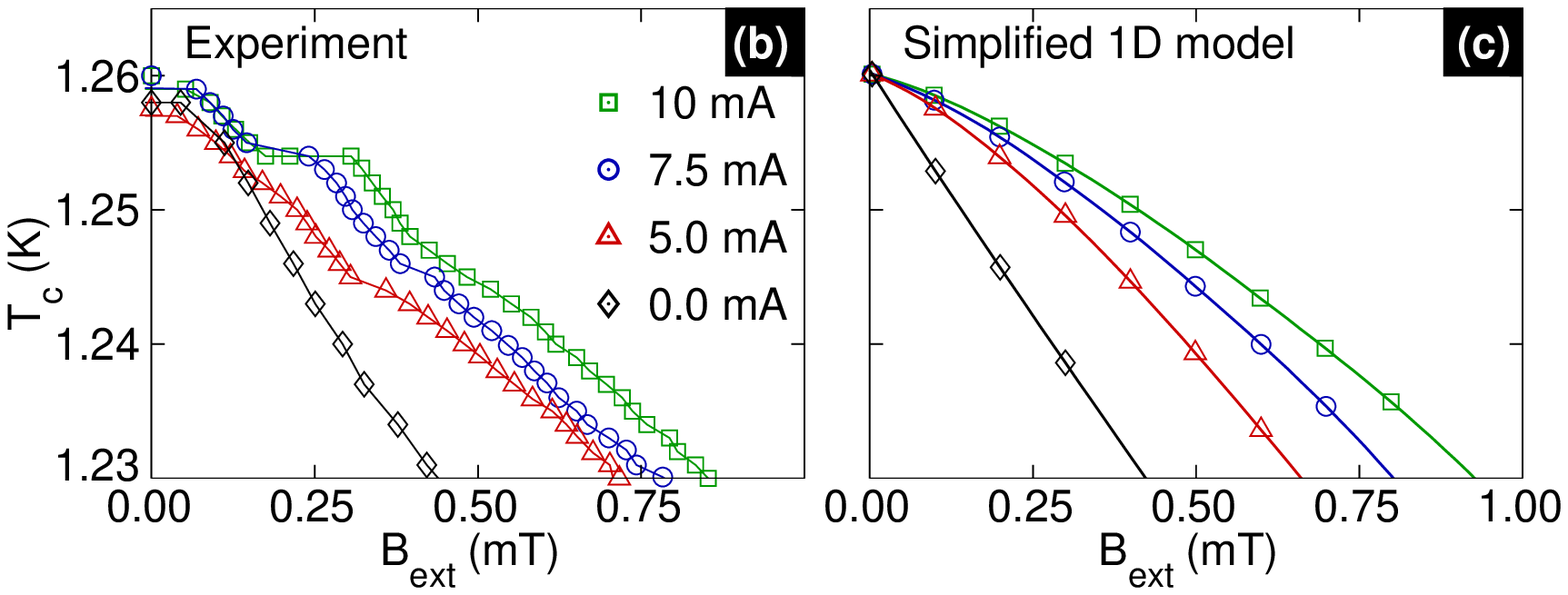}
    \caption{ (Color online) (a) Phase boundaries $T_c(B_{ext})$, composed using a 99\% criterion of the
    normal state resistance of the Al bridge, for different current
    values in the Nb wire; the zoomed parts of these curves within the
    dashed box are shown in (b). The calculated phase boundaries $T_c(B_{ext})$ for the same
    current values  $I_w=0$, 5, 7.5 and 10 mA are shown in  (c).  For
    $I_w=0$ we used Eq.~(1), while Eq.~(\ref{SimpleModel}) was used
    for nonzero $I_w$ values.
    \label{Fig2-new}}
    \end{center}
    \end{figure}

The  normal (N) -- superconductor (S) phase boundaries of the Al
strip, measured in the perpendicular external magnetic field
$B_{ext}$ using the outer voltage contacts, are shown in
Fig.~\ref{Fig2-new}(a) for different currents in the Nb wire,
$I_w$, while sending a bias current density of \mbox{$7.5\times
10^3\,$A/cm$^{2}$ through the Al strip}. To determine the
superconducting transition temperature $T_c$ a 99\% criterion of
the normal state resistance $R_n$ was used.

When the control current is zero ($I_w=0$) and thus the magnetic
field is uniform, the resulting phase boundary shows the expected
linear dependence of $T_c$ on $B_{ext}$ [black circles in
Fig.~\ref{Fig2-new}(a)]. Due to a rather high surface to volume
ratio of our mesoscopic sample and the high criterion for the
determination of $T_{c}$, it is natural to attribute the phase
boundary to the appearance of surface superconductivity
\cite{Saint-James-69}:
    \begin{eqnarray}
    T_{c3}\simeq T_{c0}\,\left\{1-0.59\,\frac{|B_{ext}|}{B_{c2}^{(0)}}\right\}.
    \end{eqnarray}
Applying this equation to the phase transition line
$T_c(B_{ext})$, measured for $I_w=0$, we determine the
superconducting critical temperature in zero field $T_{c0}\simeq
1.26~$K, as well as extrapolated to $T=0$ the coherence length
$\xi_0\simeq 175$~nm and the upper critical field
$B_{c2}^{(0)}=\Phi_0/[2\pi\xi_0^2]\simeq 10.4$~mT,
$\Phi_0=2.07\times 10^{-15}$~Wb is the magnetic flux quantum.

Interestingly, when the non-uniform component of the magnetic
field becomes nonzero ($I_w\neq 0$), the phase boundaries exhibit
a clear enhancement of the critical field for the whole
temperature range as shown in Fig.~\ref{Fig2-new}(a). This
enhancement becomes more pronounced as $I_w$ increases. Such
behavior was observed for various S/F hybrids (see, e.g., review
\cite{Aladyshkin-SUST-09} and references therein). The described
``magnetic bias" is commonly explained in terms of the local
compensation of the applied magnetic field due to nonuniform
component of the field ($B_{z}^{min}=B_{ext}-B_0$ if
$B_{ext}>B_0$) and trapping of the OP wave function at the
locations near the $B_z$ minima.

%

To illustrate the evolution of the superconducting properties in
the considered system upon varying $T$, $B_{ext}$ and $I_w$, we
performed numerical simulations within the two-dimensional (2D)
time-dependent Ginzburg-Landau (TDGL) model \cite{Comment}. For
simplicity we assume that the effect of the superfluid currents
${\bf j}_s$ on the magnetic field distribution is negligible and
consider the internal magnetic field ${\bf B}$ equal to the field
of the external sources ${\bf B}_{ext}+{\bf b}_w(x)$. This
assumptions seems to be valid (i) for mesoscopic thin-film
superconductors with lateral dimensions smaller than the effective
magnetic penetration depth $\Lambda=\lambda^2/d$ ($\lambda$ is the
London penetration depth, $d$ is the film thickness); (ii) for
superconductors at large $H$ and/or $T$ (i.e. close to the phase
transition line), when the superfluid density tends to zero. In
particular, the TDGL equations take the form
    \begin{equation}
    u \left(\frac{\partial}{\partial t}+ i\varphi \right) \psi = \tau\,\left(\psi-|\psi|^2\psi\right) + \left(\nabla + i{\bf A}\right)^2\psi,
    \label{eq:GL-1}
    \end{equation}
    \begin{equation}
    \nabla^2 \varphi = {\rm div}\, {\bf j}_s, \,{\bf j}_s = -\frac{i}{2}\tau\,\Big\{\psi^*\left(\nabla + i{\bf A}\right)\psi - \mbox{c.c.}\Big\},
    \label{eq:GL-3}
    \end{equation}
where $\psi$ is the normalized OP wave function, $\varphi$ is the
dimensionless electrical potential \cite{Comment2}, ${\bf A}$ is
the vector potential [${\rm rot\,}{\bf A}=(B_{ext}+b_{w,z})\,\,\bf
z_0$], $T$ is temperature, $\tau=(1 - T/T_{c0})$,  $u$ is the rate
of the OP relaxation, c.c. stands for complex conjugate. For the
interfaces superconductor/vacuum or superconductor/insulator the
boundary condition for $\psi$ and $\varphi$ has the standard form
    \begin{equation}
    \left(\frac{\partial}{\partial \bf n} + i A_n\right)_{\Gamma} \psi = 0,
    \quad \left(\frac{\partial \varphi}{\partial \bf n} \right)_{\Gamma} = 0,
    \label{eq:BoundCond-1}
    \end{equation}
where $\bf n$ is the normal vector to the sample's boundary
$\Gamma$.

The self-consistent TDGL modelling for the superconducting sample
with realistic dimensions (close to the experimental ones) is
impossible because of enormous data flow. Therefore, we have
restricted our consideration and analysis to a mesoscopic
superconducting rectangle: length $L=140\,\xi_0$ and width
$W=20\,\xi_0$ (e.g., if $\xi_0\simeq 0.175~\mu$m for thin-film Al
superconductors, then $L\simeq 24.5\,\mu$m and $W\simeq
3.5\,\mu$m). Since the superconducting OP wave function always has
maxima near the corners of the sample
\cite{Houghton-PL-65,Schweigert-PRB-99,Chibotaru-JMP-05}, we
formally have to assign $\psi=0$ at $x=\pm L/2$ (both ``left" and
``right" edges). This simple technical trick guarantees that the
finite length of the sample and 90$^{\circ}$-corners do not
strongly affect those OP solutions, which are localized near the
control wire and thus of main interest in the current study.

It should be noted that the field induced by the wire is
antisymmetric with respect to its middle line $x=0$,
$b_{w,z}(x,y)=-b_{w,z}(-x,y)$, therefore (i) the effective field
compensation favorable for the OP nucleation occurs at positive
$x$ values (i.e., on the right side of the wire) for $B_{ext}>0$
and vice versa; (ii) $T_c(B_{ext})=T_c(-B_{ext})$. It allows us to
consider only positive $B_{ext}$ values without loss of generality.

    \begin{figure}[h!]
    \begin{center}
    \epsfxsize=95mm \epsfbox{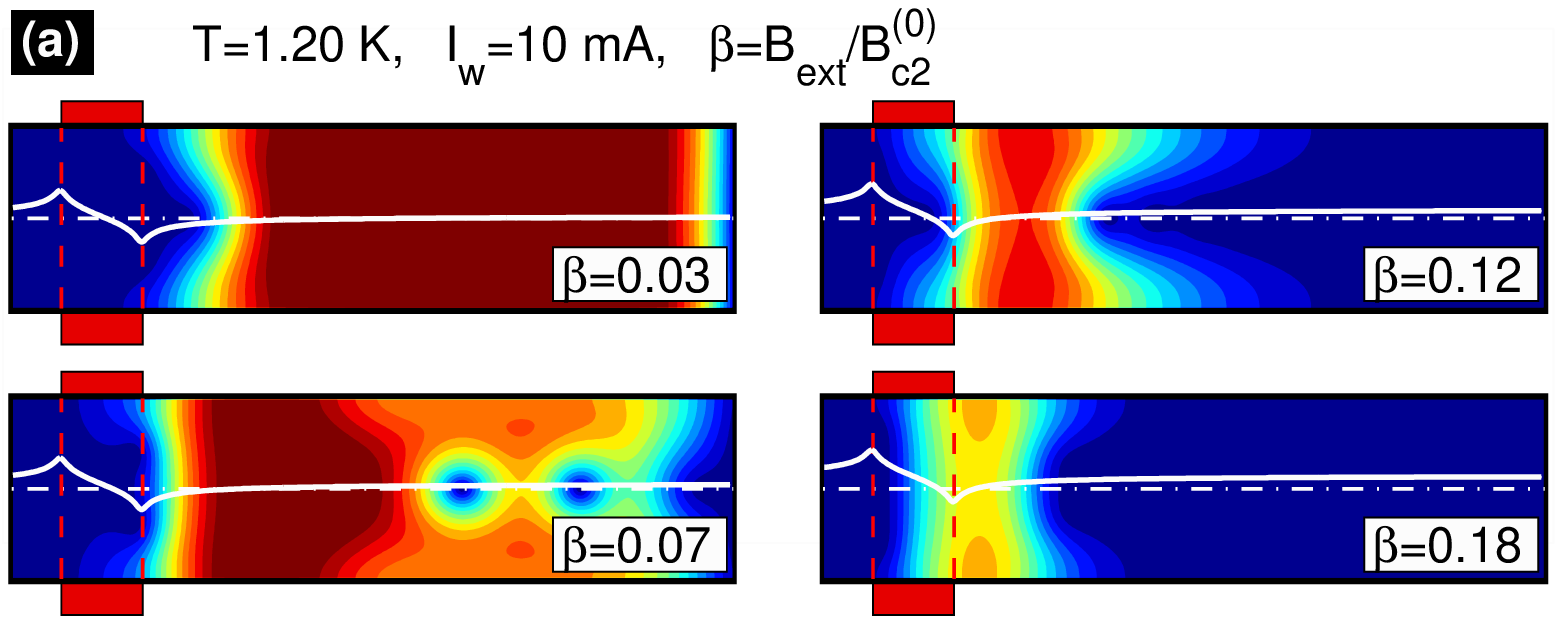}
    \epsfxsize=95mm \epsfbox{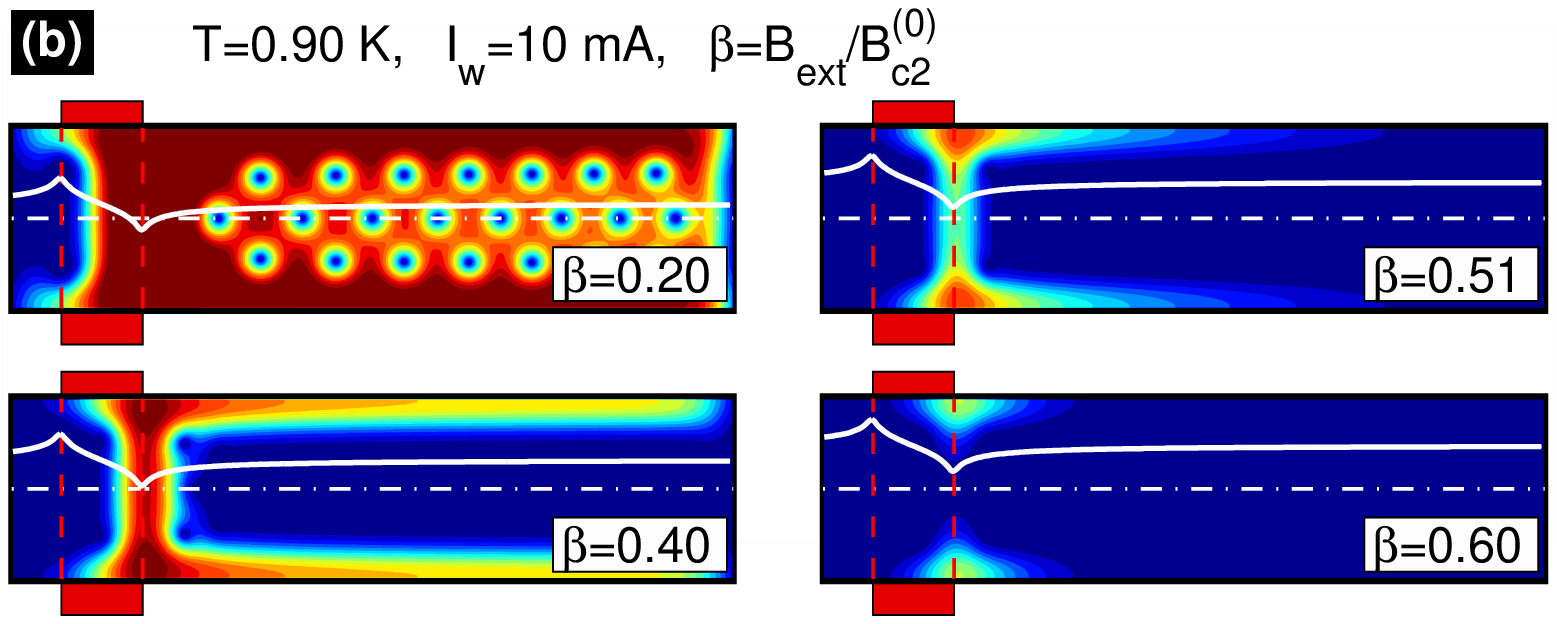}
    \epsfxsize=95mm \epsfbox{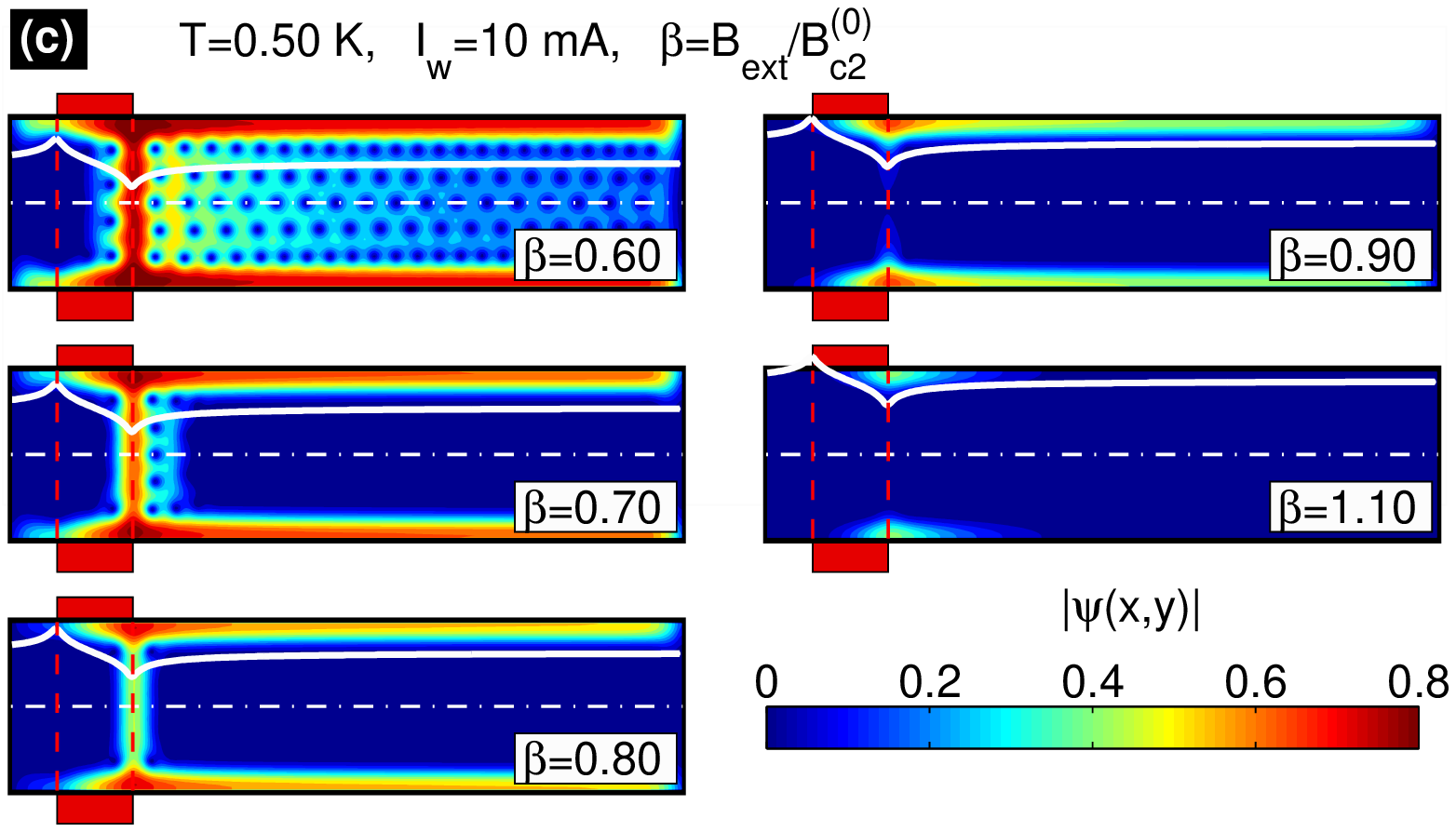}
    \caption{ (Color online) Stationary OP wave functions
    $|\psi(x,y)|$ forming in the superconducting rectangle (length 140\,$\xi_0$ and width 20\,$\xi_0$)
    in the presence of a current-carrying wire
    (current $I_w=10\,$mA, width 9$\,\xi_0$). Panel (a) corresponds to $T=1.20\,$K;
    panels (b) and (c) are for $T=0.90\,$K and $T=0.50\,$K, correspondingly;
    the $B_{ext}$ values are indicated in the plots.
    We show only the area to the right of the wire
    ($80\,\xi_0\times 20\,\xi_0$ in size) in order to better visualize
    the localized superconducting states trapped at $x>0$.
    The dash vertical lines correspond to the projections of the edges of the control wire.
    White solid lines show the resulting profile of the magnetic field $B_z(x)=B_{ext}+b_{w,z}(x)$.
    The condition $\psi=0$ at the left and right edges was used
    for the TDGL modelling, Eqs. (2)--(3).
    \label{Fig3-new}}
    \end{center}
    \end{figure}

    \begin{figure}[h!]
    \begin{center}
    \epsfxsize=95mm \epsfbox{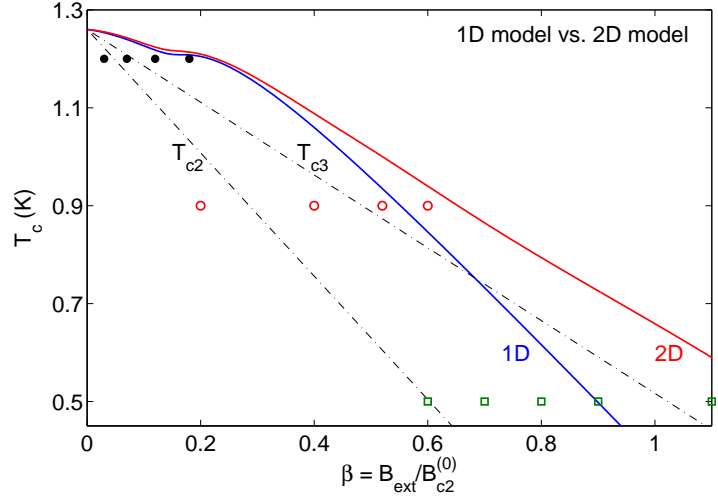}
    \caption{ (Color online) Phase boundaries calculated for the superconducting rectangle
    (length 140\,$\xi_0$ and width 20\,$\xi_0$)
    in the presence of the current-carrying wire
    (current $I_w=10\,$mA, width 9~$\xi_0$)
    within 2D TDGL model [Eqs. (2)--(3)], and
    1D linearized GL model [Eq. (5)]. We assume the identical parameters for both models (lateral dimensions,
    boundary conditions, vector potential distribution, grid size etc).
    The two dashed lines represent the phase transition line for
    the development of bulk superconductivity $T_{c2}=T_{c0}\,(1-|\beta|)$
    and for the critical field of surface superconductivity
    $T_{c3}=T_{c0}\,(1-0.59\,|\beta|)$, where $\beta\equiv B_{ext}/B_{c2}^{(0)}$. The symbols show the
    corresponding points for the images in Fig.~{\ref{Fig3-new}}.
    \label{Fig4-new}}
    \end{center}
    \end{figure}

Figure ~\ref{Fig3-new} shows typical stationary OP distributions
obtained from the TDGL model for high, intermediate and low
temperatures ($1.20$~K, 0.90~K and 0.50~K).

For finite $I_w$ and small $B_{ext}$ the OP wave function
expectedly nucleates far away from the control wire (see panel
$\beta=0.03$ in Fig.~\ref{Fig3-new}(a), $\beta\equiv
B_{ext}/B_{c2}^{(0)}$), i.e in the area where the field induced by
the wire vanishes and $T_c$ tends to $T_{c0}$ as $|B_{ext}|\to 0$.
At the same moment the OP nucleation near the middle line of the
wire, where also $B_z(x,y)\approx 0$, is less energetically
favorable due to a larger field gradient $dB_z/dx$ at $x=0$. The
depletion of bulk superconductivity at very large distances from
the wire by the applied field $B_{ext}$ corresponds to a threshold
value $\beta_0=1-T/T_{c0}\simeq 0.048$ for $T=1.20~$K. The local
suppression of the superconducting condensate (blue spots) at
rather large distances from the wire for $\beta = 0.07$ indicates
the formation of vortices by the combined magnetic field of the
control Nb wire and the applied magnetic field. For larger
$B_{ext}$ the OP wave function becomes more localized along the
$x-$axis and trapped near the region where $B_{ext}+b_{w,z}(x)=0$,
remaining more or less uniform across the Al strip (panels
$\beta=0.12$ and $\beta=0.18$). Thus, for rather high $T$ (close
to $T_{c0}$) and low $B_{ext}$, the localized OP solution moves
towards the Nb wire as $|B_{ext}|$ gradually increases, until it
finally reaches the edge of the current-carrying wire. This
process is accompanied by a monotonous decrease in $T_c$ (see
Figs. \ref{Fig2-new} and \ref{Fig4-new}), which is a direct
consequence of a shrinkage of the typical width of the OP solution
in the $x-$direction as $|B_{ext}|$ increases (an analog of the
quantum-size effect for the Cooper pairs in the nonuniform
magnetic field \cite{Aladyshkin-SUST-09}).



For lower $T$, a completely different evolution of the
superconducting properties is observed, when the inhomogeneity of
the OP wave functions across the strip becomes crucial. Indeed,
the formation of the localized superconducting state near the
minimum of the total field, $B_{z}^{min}\approx B_{ext}-B_0$,
leads to the following asymptotic behavior at $|B_{ext}|\gg B_0$:
$T^*_{c2}\simeq T_{c0}\times (1-|B_{ext}-B_0|/B_{c2}^{(0)})$. In
addition, a survival of the surface superconductivity at large
distances, where $B_z\simeq B_{ext}$, can be estimated according
Eq.~(1). Due to the different slopes $dT^{*}_{c2}/dB_{ext}$ and
$dT_{c3}/dB_{ext}$ and the different offsets, we get the point
$B_{ext}^*\approx 2.44\,B_0$, where both critical temperatures are
equal: $T^*\simeq T_{c0}\,(1-1.44 \,B_0/B_{c2}^{(0)})$ (the
similar argumentation was presented, e.g., in
\cite{Aladyshkin-PRB-07}). Substituting $B_0\simeq 0.36\,
B_{c2}^{(0)}$ (estimated for $I_w=10\,$mA), we get the threshold
temperature $T^*\simeq 0.6\,$K. It means that for intermediate
temperatures ($T=0.90\,$K, panel (b) in Fig.~\ref{Fig3-new}) the
gradual increase in $B_{ext}$ causes subsequently the development
of bulk superconducting state (the plot labelled $\beta=0.20$),
the complete suppression of bulk superconductivity ($\beta=0.40$),
and then the suppression of surface (or edge-assisted)
superconductivity at large distances from the wire and the
survival of the state localized near the right edge of the wire
($\beta=0.51$), since $T_{c2}^*>T_{c3}$. Finally,
superconductivity survives in a form of 2D patterns localized in
both directions and centered at the points where the wire's right
edge intersects the superconducting sample ($\beta=0.60$). In
contrast to that, for low temperatures (at $T=0.50\,$K, panel (c)
in Fig.~\ref{Fig3-new}) the enhanced superconductivity along the
wire's edge is suppressed before the destruction of edge-assisted
superconductivity (images $\beta=0.80$ and $\beta=0.90$), since
$T_{c2}^*<T_{c3}$ in this temperature range.

%
%
%

%
%
%
%
%
%
%
%
%


To show better the observed difference in the shape of localized
OP patterns at high and low temperatures we calculate numerically
the phase transition lines $T_c(B_{ext})$ using the described 2D
TDGL model, Eqs. (2)--(4) and 1D linearized Ginzburg--Landau (GL)
equation
    \begin{equation}
    \tau\,\psi + \left(\nabla + i{\bf A}\right)^2\psi = 0,
    \end{equation}
assuming $\psi=\psi(x,y)$ and $\psi=\psi(x)$, respectively, for
the same patterns $B_z$ and ${\bf A}$ and the same boundary
condition $\psi=0$ at $x=\pm L/2$. The used 1D model was described
in detail in \cite{Aladyshkin-PRB-06}. It is easy to see that for
low $B_{ext}$ and high $T$ both models give almost identical
results, what supports our conclusion that the appearing OP
solutions are almost uniform  across the strip and the OP
inhomogeneity in the $y-$direction can be disregarded. For
intermediate and low temperatures and  for large $B_{ext}$ the
phase boundaries $T_c(B_{ext})$ reveal a linear behavior with
different slopes: $dT_c/dB_{ext}\simeq dT_{c2}/dB_{ext}$ for 1D
model and $dT_c/dB_{ext}\simeq  0.59\,dT_{c2}/dB_{ext}$ for 2D
model. It indicates that in the limit $|B_{ext}|\gg B_0$ the
superconductivity is trapped both near the strip edges (in order
to correspond to the slope typical for the surface
superconductivity) and at the minimum of the local magnetic field
(in order to explain the parallel shift of the high-field
asymptote of $T_c(B_{ext})$ in higher field).

Since we are mainly interested in the migration of the OP along
the Al strip in low magnetic fields, we can propose a very simple
description based on 1D linearized GL model, neglecting the
finiteness of the Al strip in the $y-$direction
\cite{Aladyshkin-PRB-03}. If superconductivity is confined within
an area where the local magnetic field can be approximated by a
linear dependence $B_z(x)=b^{\prime}_{w,z}(x_0)\,(x-x_0)$, where
$x_0$ is the point of zero total magnetic field
($B_{ext}+b_{w,z}(x_0)=0$), for the phase boundary we obtain as a
rough estimate
    \begin{eqnarray}
    T_{c}\simeq T_{c0}\,\left\{1-\pi^{2/3}\, \xi_0^{2/3}\left[b^{\prime}_{w,z}\right]^{2/3}\Phi_0^{-2/3}\right\}.
    \end{eqnarray}
Considering the generic case of a non-uniform field $b_{w,z}=\mu_0
I_w/(2\pi x)$, induced by an infinitely thin cylindrical wire
carrying a control current $I_w$, at the large distances from the
wire one gets $|b^{\prime}_{w,z}(x_0)|=2\pi B_{ext}^2/(\mu_0 I_w)$
and hence
    \begin{eqnarray}
    T_{c}\simeq T_{c0}\,\left\{1-\xi_0^{2/3} \left(\frac{\pi B_{c2}^{(0)}}{\mu_0 I_w}\right)^{2/3}
    \left(\frac{B_{ext}}{B_{c2}^{(0)}}\right)^{4/3} \right\},
    \label{SimpleModel}
    \end{eqnarray}
which seems to be valid only for low fields, $|B_{ext}|\ll B_0$,
when the OP wave function is located far from the current-carrying
wire. We compare the experimental data, obtained for low $B_{ext}$
[Fig.~\ref{Fig2-new}(b)], with the results of this model
[Fig.~\ref{Fig2-new}(c)], where we use the same parameters
$T_{c0}$, $\xi_0$ and $B_{c2}^{(0)}$. We conclude that the 1D
model works quite well for describing the $T_c$ suppression and
the OP migration along the Al strip upon sweeping $B_{ext}$. By
changing both $B_{ext}$ and $I_w$ we directly verify that the
nucleation of superconductivity is controlled not only by the
local magnetic field, but also by the gradient of the field in the
area where superconductivity is confined.
    \begin{figure}[h!]
    \begin{center}
    \epsfxsize=95mm \epsfbox{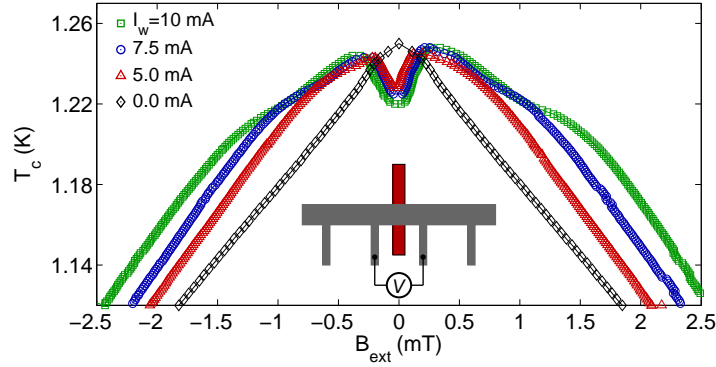}
    \caption{ (Color online) N/S
    phase boundaries determined using a 95\% criterion of the normal
    state resistance of the Al film for different current values in
    the Nb wire, measured at the inner contacts of the bridge.
    \label{Fig5-new}}
    \end{center}
    \end{figure}

    \begin{figure}[h!]
    \begin{center}
    \epsfxsize=95mm \epsfbox{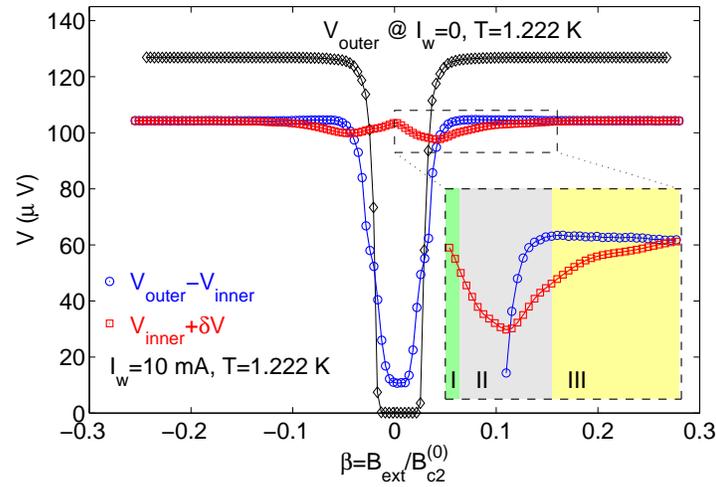}
    \caption{ (Color online) Typical dependences $V$ on $B_{ext}$ measured for both outer and
    inner voltage contacts at temperatures close to $T_{c0}$,
    when one-dimensional OP migration along the Al wire is anticipated.
    It should be noted that normal state resistances $R^{(n)}_{outer}=V^{(n)}_{outer}/I_0\simeq 0.45\,\Omega$ and
    $R^{(n)}_{inner}=V^{(n)}_{inner}/I_0\simeq 2.53\,\Omega$,
    measured at the corresponding contacts at high magnetic field, differ
    approximately 5.6 times due to the difference in length between these contacts ($I_0=50\,\mu$A is the bias current).
    Therefore, for a better visualization we show the difference
    $V_{outer}-V_{inner}$ and $V_{inner}+\delta V$, which can be attributed exclusively to the voltage
    drop {\it outside} the inner contacts and {\it between} the inner
    contacts,respectively.
    The offset $\delta V\simeq 82\,\mu$V was introduced
    in order to equalize the position of the plateaus on $V_{outer}-V_{inner}$ and $V_{inner}+\delta V$
    at large $B_{ext}$.
    \label{Fig6-new}}
    \end{center}
    \end{figure}

It is worth noting that this magnetic field profile is very
similar to the field produced by a single domain wall in a thin
ferromagnetic layer with out-of-plane magnetization
\cite{Aladyshkin-PRB-03}. The N-S phase boundary of a
superconducting film on top of such a ferromagnetic system
calculated in \cite{Aladyshkin-PRB-03} looks very similar to that
observed in our experiments. As a result, we can claim that our
S/Em hybrid system behaves as a ferromagnetic film with a pinned
straight domain wall with tunable saturated magnetization
underneath a superconducting thin film, i.e. a situation, which
cannot be easily achieved in the S/F bilayers.

Up till now we presented the superconducting properties measured
with the outer voltage contacts located at the distance 50$\,\mu$m
from the Nb wire. To prove the concept of the OP localization
directly, we prepared a second pair of inner voltage contacts
closer to the Nb wire (10$\,\mu$m from the wire). The results of
these measurements are shown in Fig.~\ref{Fig5-new}. In this case
to determine the phase boundary $T_c(B_{ext})$ we used a 95\%
criterion, so that the same condition for detection of
superconductivity as the 99\% criterion for the 100$\,\mu$m
voltage pad separation is obtained for the inner contacts of
distance 20$\,\mu$m.

The key finding is that $T_c$ measured by the inner contacts is
always lower than $T_c$ for the outer contacts, provided
$|B_{ext}|$ is rather small (up to 0.5 mT). Nevertheless the
high-field asymptotes $T_c(B_{ext})$ for both types of contact
arrangements expectedly coincides for high fields (compare
Fig.~\ref{Fig2-new}(a) and Fig.~\ref{Fig5-new}). This implies that
the inhomogeneous superconductivity is located within the inner
contacts for high fields and outside these contacts for low
fields. Indeed, in our magnetoresistive measurements using the
inner contacts at rather high temperatures and $B_{ext}\approx 0$
we cannot detect superconductivity nucleating outside these
contacts. Therefore the measured $T_c$ must be lower than the
critical temperature for the OP solution localized far away from
the Nb wire. This is convincing experimental evidence for the
field-dependent OP localization in the non-uniform magnetic field.
We observed that upon increasing $B_{ext}$, the localized
superconductivity shifts toward the wire, leading to a
non-monotonous variation of the resistance measured between the
inner contacts and, correspondingly, to the non-monotonous
variation in $T_c$ (see Fig.~\ref{Fig6-new}). Unlike the reentrant
dependence $V_{inner}$ vs. $B_{ext}$, the voltage drop measured at
the outer contacts monotonously increases as $B_{ext}$ increases,
since localized superconductivity always nucleates between the
outer contacts and never leaves this area. Considering the inset
in Fig.~\ref{Fig6-new}, we propose the following interpretation of
our findings: in the region I the voltage drop
$V_{outer}-V_{inner}$, attributed to the area between the outer
and inner contacts solely, is minimal and almost independent on
$B_{ext}$, while $V_{inner}$ is maximal, therefore there is no
global superconductivity at this temperature in the sample and the
OP wave function has to be localized between the outer and inner
contacts. In the region II both $V_{outer}-V_{inner}$ and
$V_{inner}$ strongly depend on $B_{ext}$, and as a result, the OP
wave function should be located somewhere in the vicinity of the
inner contacts. Finally, in the region III the difference
$V_{outer}-V_{inner}$ becomes equal to the field-independent value
corresponding to the normal state and $V_{inner}$ is still smaller
than its normal value, therefore the OP wave function is
definitely trapped between the inner contacts (i.e., near the
control wire). This behavior is in good agreement with the
theoretical predictions.

The position of the $T_c$ maximum for the inner contacts
measurements can be attributed to the effective compensation of
the built-in magnetic field at the inner voltage contacts. Due to
the small asymmetry in the position of the voltage contacts at the
opposite sides of the wire as a result of fabrication
imperfections, both peaks occur at different fields:
$B_{ext}=-0.32\,$mT and $B_{ext}=0.26\,$ mT, which is on the order
of what we expect from the induced field at the voltage contacts
for a current of 10 mA. More interestingly, the $T_c$'s
corresponding to these peaks are slightly different, clearly
showing the influence of the field gradient at the point of
localization on $T_c$.

Summing up, we have studied the OP localization in an Al strip
subjected to an inhomogeneous field with tunable intensity induced
by a current-carrying wire. The OP migration along the strip upon
varying $B_{ext}$ and $I_w$ has been detected by using multiple
voltage contacts. We have shown that the critical temperature at
the compensated positions is dependent on the local variation of
the magnetic field. Interestingly, we demonstrate that both
reentrant and non-reentrant superconducting phase boundaries can
be obtained depending on where the voltage drop is recorded.

This work was supported by Methusalem Funding by the Flemish
Government, and the FWO, the Carl-Zeiss-Stiftung, and the Deutsche
Forschungsgemeinschaft (DFG) via the SFB/TRR 21, the Russian Fund
for Basic Research, RAS under the Program "Quantum physics of
condensed matter" and FTP "Scientific and educational personnel of
innovative Russia in 2009-2013". W.G., J.V.d.V and A.V.S. are
grateful for the support from the FWO-Vlaanderen.


\begin{thebibliography}{30}
\bibitem{Comment}
The described simulations were performed using the
Windows-oriented solver GLDD, developed in the Institute for
Physics of Microstructures RAS.
\bibitem{Comment2}
We use the following units: $m^*\sigma_n
\beta/(2e^2\tilde{\alpha})$ for time, the coherence length $\xi_0$
at temperature $T=0$ for distances, $\Phi_0/(2\pi\xi_0)$ for the
vector potential; $\hbar e |\tilde{\alpha}|/(m^*\sigma_n \beta)$
for the electrical potential, and
$4e\tilde{\alpha}^2\xi_0/(\hbar\beta)$ for the current density,
where $\alpha=-\tilde{\alpha}\,\tau$ and $\beta$ are the
conventional parameters of the GL expansion, $e$ and $m^*$ are
charge and the effective mass of carriers, $\sigma_n$ is the
normal state conductivity.
\end{thebibliography}
\end{document}